\documentclass[onecolumn,noshowpacs,preprintnumbers,amsmath,amssymb]{revtex4}
\usepackage{amsfonts}
\usepackage{amsmath}
\usepackage{graphicx}
\def \ee{\end{equation}}
\def \be{\begin{equation}}
\def \bea{\begin{eqnarray}}
\def \eea{\end{eqnarray}}
\def\lsim{\lower.5ex\hbox{$\; \buildrel < \over \sim \;$}}
\def\gsim{\lower.5ex\hbox{$\; \buildrel > \over \sim \;$}}

\def\bra#1{\mathinner{\langle{#1}|}}
\def\ket#1{\mathinner{|{#1}\rangle}}

{\catcode`\|=\active\gdef\Braket#1{\left<\mathcode`\|"8000\let|\bravert {#1}\right>}}
\def\bravert{\egroup\,\vrule\,\bgroup}
\begin{document}

\title{Exploring a Tractable Lagrangian for Arbitrary Spin}

\author{Benjamin Koch and Nicol\'as Rojas}
\affiliation{Pontificia Universidad Cat\'olica de Chile.\\ Av. Vicu\~na Mackenna
4860. Macul.\\ Santiago de Chile.}

\date{\today}
\email{bkoch@fis.puc.cl}

\begin{abstract}
In this student project, performed at the
Pontificia Universidad Catolica de Chile in 2011,
a simple Lagrangian is proposed
that by the choice of the representation
of SU(2), gives rise to field equations
for arbitrary spin.
In explicit examples it is shown, how the
Klein-Gordon, the Dirac, and the
Proca equation can be obtained from this Lagrangian.
On the same footing, field equations for arbitrary spin are
given. Finally, symmetries are discussed, the fields are quantized,
their statistics is deduced, Feynman rules are derived,
and problems of the formulation are discussed.
\end{abstract}

\maketitle
\section{Introduction}
The program of formulating field equations
for arbitrary spin was started by Dirac, Pauli, and
Fierz \cite{Dirac:1936tg,Fierz:1939ix}.
Since then it has been investigated from
various perspectives, leading to a big variety
of possible formulations and applications
\cite{Gelfand:1948,Bargmann:1948ck,Weinberg:1964cn,Chang:1967zz,Tung:1967zz,Hagen:1970wn,
Hurley:1971nz,Hurley:1972ju,Singh:1974qz,Singh:1974rc,
Gershun:1979fb,Berends:1985xx,Siegel:1986de,Siegel:1986zi,
Howe:1988ft,Siegel:1988yz,Berkovits:1996tn,Metsaev:1997nj,Francia:2002aa,Niederle:2004bw,
Vasiliev:2004qz,Savvidy:2005vm,Bekaert:2006us,Francia:2007ee,Francia:2008ac,Engquist:2008rt,
Campoleoni:2008jq,Bengtsson:2009nk,Buchbinder:2009pa,Manvelyan:2010jr,
Campoleoni:2011hg,Polyakov:2011sm,Chicherin:2011sm, Montero:2011za}.
In parallel to the growing number of formulations,
conditions such as external field interactions,
subluminal propagation,
and curved background were studied that would allow to
prefer some formulations and reject others
\cite{Buchdahl:1958xv,Velo:1970ur,Buchdahl:1982ni,Illge:1999tb,Illge:1986vs,
Illge:1993cd,Deser:2001dt,Sorokin:2004ie,Zecca:2007ab}.
One tends to believe that a self consistent interacting Lagrangian for
arbitrary spin, would be of great interest to
young students and
researchers entering the particle physics community.
However, due to their level of complexity and mathematical abstraction
such formulations gained less attention than 
one might have expected \cite{Belinicher:1974am,Green:1978dz}.
Many of the recent approaches to arbitrary spin 
can be categorized \cite{Buchbinder:2007ak,Buchbinder:2008ss}
into the BRST type of approach and the geometric approach.
In order to keep the objective of accessibility and simplicity
no such formal construction is intended in this paper.

The aim of this summer project was to approach
this complex topic in an independent and for young
students tractable way.
This is done by introducing and exploring
a simple toy Lagrangian for arbitrary spin.
This ad hoc formulation
that was chosen since it is a straight forward generalization
of the simplest relativistic field equation: the Klein-Gordon equation.

The organization of this report is as follows:
First, the Lagrangian is introduced 
in an abelian and in a non-abelian gauge formulation.
Second, it is explicitly shown how 
this Lagrangian gives rise
to the Klein Gordon equation, the Dirac 
equation, and the Proca equations. Then 
the general equations 
of motion are given in a form that is equivalent
to an already established formulation of 
the arbitrary spin equations.
Third, symmetries and conserved quantities
of this Lagrangian are explored. 
Fourth, field quantization 
in this approach is discussed and a surprising
statistics for those quanta is found.
Finally, Feynman 
rules are derived. \\


\section{The Lagrangian}
\label{sectlag}
The starting point is a Lagrangian 
that has derivative terms of second order
\be\label{Lmaster}
{\mathcal{L}}_{s}=
 (D_\mu \Omega)^\dag D^\mu \psi-i g_s e\Omega^\dag H^s_{\mu \nu}
F^{\mu \nu}\psi -m^2\Omega^\dag \psi+c.c.\,.
\ee
It contains interactions with an external gauge
potential $A^\mu$ and the corresponding field strength tensor
$F^{\mu \nu}=\partial^\mu A^\nu-\partial^\nu A^\mu$.
The covariant derivative is $D_\mu=\partial_\mu-ieA_\mu$.
The spin dependent g-factor $g_s$ 
is chosen such that it accommodates the
inverse proportionality that was
found by \cite{Belinfante:1953zz,Hagen:1970wn}
\be
g_s=\left\{
\begin{array}{ccc}
 1  & {\mbox{for}} &s=0\\
\frac{1}{4s} &{\mbox{for}}& s>0
\end{array}
\right.\quad. \label{gs}
\ee
$\Omega$ and $\psi$ are matter fields 
of an a priory undefined spin and they
are not necessarily independent.
The linear operator $H^s_{\mu \nu}$ is defined as
\be\label{Htensor}
H^s_{0i}=-H^s_{i0}=\Sigma^s_i\;,\quad\;
H^s_{ij}=-\frac{1}{2}[\Sigma^s_i,\Sigma^s_j]\quad.
\ee
The $\Sigma^s_i$ are hermitian $(2s+1)\otimes (2s+1)$
matrices that fulfill the algebra
\be\label{su2}
\left[\Sigma^s_j,\Sigma^s_k \right]=2i\epsilon_{jkl}\Sigma^s_l\quad.
\ee
Under Lorentz transformations the operator $H^s_{\mu \nu}$
does not change, while the fields transform according to
\be\label{Ltrans}
\psi'=e^{\frac{1}{2}(-i\alpha_j-\beta_j)\Sigma^s_j}\psi\;,\quad
\Omega'=e^{\frac{1}{2}(-i\alpha_j+\beta_j)\Sigma^s_j}\Omega\quad.
\ee
Here, $\alpha_j$ is a rotation angle around the $j$ axes
and $\beta_j$ is a Lorentz boost along the $j$ direction. 
From (\ref{Ltrans}) one can read off that that $\psi$ lives
in the $(s,0)$ representation of the Lorentz group,
while $\Omega$ lives in the $(0,s)$ representation.

For some applications, it is 
convenient to note that the $2s+1$ compenent matter
fields can be combinend to a $2(2s+1)$ component matter field
\be
\Psi(x)=
\left[
\begin{tabular}{c}
 $\psi(x)$\\
 $\Omega(x)$
\end{tabular}
\right]\quad.
\ee
One further defines $(2(2s+1))$ dimensional matrices
\be\label{gamma0}
\gamma^0=\left(
\begin{tabular}{cc}
 0& $\openone$\\
 $\openone$ &0
\end{tabular}
\right)\;
, \;
\gamma^5=\left(
\begin{tabular}{cc}
 $\openone$&0\\
 0&-$\openone$
\end{tabular}
\right)\quad.
\ee
This allows to define an
adjoint field $\bar{\Psi}=\Psi^\dagger \gamma^0$ 
and the operator $P_\pm=(1\pm
\gamma^5)/2$ 
that projects $\Psi$ back
onto the fields $\psi$ and $\Omega$.
Doing this, the Lagrangian (\ref{Lmaster}) may also 
be rewritten as
\begin{eqnarray}\label{Lmaster2}
\mathcal{L}_{a.} &=& D_{\mu}^{\dagger}\bar{\Psi}D^{\mu}\Psi -
ig_s e\bar{\Psi}\mathcal{H}_{(s)}^{\mu\nu}F_{\mu\nu} \Psi -
m^2\bar{\Psi}\Psi \quad.
\end{eqnarray}
In the same way, one may write the Lagrangian in 
with nonabelian gauge symmetry as
\begin{eqnarray}\label{Lmaster22}
\mathcal{L}_{n.a.} &=& D_{\mu}^{\dagger}\bar{\Psi}D^{\mu}\Psi -
ig_s e\bar{\Psi}\mathcal{H}_{(s)}^{\mu\nu}F_{\mu\nu}^a T_a \Psi -
m^2\bar{\Psi}\Psi \quad.
\end{eqnarray}
The covariant derivative has the form:
\begin{eqnarray}
D_\mu \Psi &=& \partial_\mu \Psi -ieA_{\mu}^a \left( \begin{array}{cc} 
                              T_a  & 0 \\ 
                              0 & T_a \end{array} \right) \Psi \quad,\\
D_\mu^{\dagger} \bar{\Psi} &=& \partial_\mu \bar{\Psi} + ieA_{\mu}^a
\bar{\Psi}\left( \begin{array}{cc} 
                              T_a  & 0 \\ 
                              0 & T_a \end{array} \right)\quad,
\end{eqnarray}
and the field strength tensor is
\begin{eqnarray}
F_{\mu\nu}^a &=& \partial_\mu A^a_{\nu} - \partial_\nu A^a_{\mu} + ieA_{\mu}^b
A_{\nu}^c f^{a}_{bc}\quad.
\end{eqnarray}
The $T_a$ are the generators of the $SU(N)$ gauge group
and the $f^{a}_{bc}$ are the structure constants 
of the same gauge group.
Note that the fields $\Omega$ and $\psi$ (therefore also
$\Psi$) have the same colour
quantum numbers. In the Lagrangian (\ref{Lmaster2}) 
the commutator of spin matrices $\Sigma_j$ was 
redefined as:
\begin{eqnarray}\label{mathcalH}
\mathcal{H}_{(s)}^{\mu\nu} = \left( \begin{array}{cc} 
                              H_{(s)}^{\mu\nu}  & 0 \\ 
                              0 & H_{(s)}'^{\mu\nu} \end{array} \right)\quad.
\end{eqnarray}
 This
matrix contains the definition
\be\label{Hptensor}
H'^{s}_{0i}=-H'^{s}_{i0}=-\Sigma^s_i\;,\quad\;
H'^{s}_{ij}=-\frac{1}{2}[\Sigma^s_i,\Sigma^s_j]\quad.
\ee
Please note that gravitational couplings are
not considered.
Since the consistent implementation of gravitational
interactions into Lagrangians with arbitrary spin
seems to be hardly possible
\cite{Buchdahl:1958xv,Velo:1970ur,Buchdahl:1982ni,Illge:1999tb,Illge:1986vs,
Illge:1993cd,Deser:2001dt,Sorokin:2004ie,Zecca:2007ab}, 
it is expected that this kind of problem will most probably also
appear in our Lagrangian.
\newpage
\section{Field equations}
In this section the action is used to derive 
field equations for the particular cases $s=0,\;1/2\;,1$.
Finally, a field equation for
arbitrary spin is derived.
\subsection{Spin zero}
%
For fields without spin ($s$=0) the representation of SU(2) is
\be
\Sigma^0_j=0\quad,
\ee
which implies $H^0_{\mu\nu}=0$. 
This was possible
due to the definition $g_s=1$, in
the equation (\ref{gs}).
The two fields are independent and their equations of
motion are complex Klein-Gordon equations
\bea\label{eomKG}
D_\mu D^\mu \psi +m^2\psi&=&0\quad,\\ \nonumber
D_\mu D^\mu \Omega +m^2\Omega&=&0\quad.
\eea
Please note
that one might rotate the two fields $\psi$ and $\Omega$
in the Lagrangian by $\pi/2$ and find that one of the resulting fields is
actually a ghost field. This observation will
confirmed from a different and more general point of view when
the fields are quantized.
%
\subsection{Spin one half}\label{secshalf}

For spin one half one has $s=1/2$, which
is represented by the two dimensional Pauli matrices
\be\label{sigmahalf}
\Sigma^{1/2}_j=\sigma_j\quad.
\ee
The fields $\Omega$, $\psi$ are spinors with two components.
After using the definition (\ref{Htensor})
\be
H^{1/2}_{0i}=-H^{1/2}_{i0}=\sigma_i\,,\quad
H^{1/2}_{ij}=-\frac{1}{2}[\sigma_i,\sigma_j]\quad,
\ee
the Lagrangian reads
\be\label{LDir}
{\mathcal{L}}_{1/2}=
 (D_\mu \Omega)^\dag D^\mu \psi-\frac{i}{2}e\Omega^\dag 
 H^{1/2}_{\mu \nu}
F^{\mu \nu}\psi -m^2\Omega^\dag \psi+c.c.\;.
\ee
In order to rewrite this Lagrangian in a more familiar form
one defines
\begin{eqnarray}
D^+ \equiv D_0+\sigma_j D_j\equiv \bar{\sigma}^\mu D_\mu\quad,\\ \nonumber
D^-\equiv D_0-\sigma_j D_j\equiv \sigma^\mu D_\mu\quad.
\end{eqnarray}
Due to the Clifford algebra of the Pauli matrices one has
\be\label{keyshalf}
\bar{\sigma}_\mu \sigma_\nu=g_{\mu \nu}+H^{1/2}_{\mu \nu}\quad,
\ee
which allows to write the Lagrangian (\ref{LDir}) as
\be\label{LBrown}
{\mathcal{L}}_{1/2}=m^{-1}(D^+\Omega)^{\dag}(D^-\psi)-m \Omega^\dag\psi
+c.c.\quad.
\ee
According to (\ref{Ltrans}) the spin one half fields in this
Lagrangian transform under Lorentz transformations like
\bea\label{Ltranshalf}
\psi'=e^{\frac{1}{2}(-i\alpha_j-\beta_j)\sigma_j}\psi\,,\quad
\Omega'=e^{\frac{1}{2}(-i\alpha_j+\beta_j)\sigma_j}\Omega\quad.
\eea
The above Lagrangian was proposed 
and discussed in 1958 by Brown \cite{Brown:1958zz} 
for the description of spin one half. Its equivalence to
the Dirac formulation can be shown at the level of the equations
of motion.
By varying (\ref{LBrown}) with respect to the spinor fields one obtains two 
 equations of motion 
\bea\label{eomLB1}
(D^+D^- +m^2)\psi&=&0\quad,\\ 
\label{eomLB2}
(D^-D^+ +m^2)\Omega&=&0\quad.
\eea
The first one of those equations is actually the Feynman-Gell-Mann
equation \cite{Feynman:1958ty}
and the second is an anomalous Feynman-Gell-Mann equation.
Applying $D^-$ to the left hand side of (\ref{eomLB1})
and $D^+$ to the left hand side of (\ref{eomLB2}) one sees
that the two fields are related.
Instead,
$D^-\psi$ obeys the same equation of motion as $\Omega$
and $D^+\Omega$ obeys the same equation of motion as $\psi$.
In \cite{Brown:1958zz} it is shown how this fact and the condition
of an hermitian Hamiltonian motivates the field equations
\be\label{WeylEq}
iD^+\Omega=m\psi,\quad iD^-\psi=m\Omega \quad.
\ee
The two equations in (\ref{WeylEq})
can be combined to a single equation for
a four component spinor
\be
(D_0-\alpha_i D_i-\beta m)
\left(\begin{array}{c}
 \psi\\
\Omega
\end{array}\right)=0 \quad,
\ee
with
\be
\alpha_i=
\left(
\begin{array}{cc}
 \sigma_i & 0\\
0 & -\sigma_i
\end{array}
\right),\quad
\beta=
\left(
\begin{array}{cc}
 0 & \openone \\
\openone & 0
\end{array}
\right)\quad.
\ee
By multiplying this with $\beta$
from the left hand side one obtains the
conventional form of the Dirac 
equation for the spinor
$\Psi=(\psi, \Omega)$
\be\label{Dirac}
(i\gamma^\mu D_\mu -m)
\Psi=0\quad,
\ee
where $\gamma_\mu=(\beta,\beta \alpha_i)$.
This proves the close connection between the Dirac equation
and the action (\ref{LDir}).
Please note that the solutions of (\ref{Dirac}) are certainly
also solutions of (\ref{eomLB1} and \ref{eomLB2}), but the inverse
statement is however not necessarily true since 
negative energy wave functions with
 $(i\gamma^\mu D_\mu +m)\Psi=0$ would also solve 
(\ref{eomLB1} and \ref{eomLB2}).

\subsection{Spin one}\label{secs1}

For $s=1$ a three dimensional representation
of the algebra (\ref{su2}) is needed.
We choose the adjoint representation
\be
(\Sigma^1_k)_{lm}=2i \epsilon_{klm}\quad.
\ee
For this representation the linear operator 
$(H^s_{\mu\nu})_{mn}$ reads
\be\label{originalH}
(H^{1}_{0k})_{mn}=2i\epsilon_{kmn},\;\,
(H^{1}_{ij})_{lm}=2(\delta_{il}\delta_{jm}-\delta_{im}\delta_{jl})\;,
\ee
where the Latin indices run from one to three.
The spin one Lagrangian is then
\be\label{LMaxwell}
{\mathcal{L}}_{1}=
 (D_\mu \Omega)^\dag D^\mu \psi-\frac{i}{4}e\Omega^\dag H^{1}_{\mu \nu}
F^{\mu \nu}\psi -m^2\Omega^\dag \psi+c.c.\;,
\ee
where $\Omega=\Omega_k$ and $\psi=\psi_k$ 
have three complex 
components. The equations of motion are
\bea\label{eoms11}
(D_\mu D^\mu+m^2) \psi_m - i e F^{0j} \epsilon_{jmn}\psi_n
+ i e F^{mn}\psi_n
&=&0 \quad,\\ \label{eoms12}
(D_\mu D^\mu+m^2) \Omega_m + i e F^{0j} \epsilon_{jmn}\Omega_n
+ i e F^{mn}\Omega_n 
&=&0\quad.
\eea
The six complex components $\psi_k$, $\Omega_k$ can be 
expressed in terms of the six complex fields $\tilde E_k$ and $\tilde B_k$
by using the transformation 
\be\label{putEB}
\psi_k=\tilde B_k+i\tilde E_k\,
,\;
\Omega_k=\tilde B_k-i\tilde E_k\,;\quad\quad
\tilde B_k=\frac{\psi_k+\Omega_k}{2}\,,\;
\tilde E_k=\frac{\psi_k-\Omega_k}{2i}\quad.
\ee
A hint for interpreting those fields
comes from their behavior under Lorentz transformations.
By expanding (\ref{Ltrans}) for infinitesimal rotations $\alpha_j$ and boosts
$\beta_j$,
one sees that $\tilde E$ transforms
like an electric field and that $\tilde B$
transforms like a magnetic field 
\bea
\tilde B'_k\approx
\tilde B_k+\alpha_j\epsilon_{jkl}\tilde B_l+\beta_j\epsilon_{jkl}
\tilde E_l\quad,\\ \nonumber
\tilde E'_k\approx
\tilde E_k+\alpha_j\epsilon_{jkl}\tilde E_l-\beta_j\epsilon_{jkl}
\tilde B_l\quad.
\eea
Please note that although they transform
in the same way under Lorentz transformations,
the fields introduced here
are not the external electric
and magnetic fields
($\tilde E_k \neq E_k$ and $\tilde B_k \neq B_k$).
With (\ref{putEB}) one can
combine the equations of motion for those fields
\bea\label{eoms13}
\frac{(\ref{eoms11})+(\ref{eoms12})}{2}\Rightarrow &
(D_\mu D^\mu+m^2)\tilde B_m - i e F^{0i} \epsilon_{imn}\tilde E_n
+ i e F^{mn}\tilde B_n
=&0 \quad,\\ \label{eoms14}
\frac{(\ref{eoms11})-(\ref{eoms12})}{2i}\Rightarrow&
(D_\mu D^\mu+m^2)\tilde E_m + i e F^{0i} \epsilon_{imn}\tilde B_n
+ i e F^{mn}\tilde E_n
=&0 \quad.
\eea
Further simplification is achieved by defining 
\be
G_{oi}=\tilde E_i,\quad G_{ij}=\epsilon_{ijk}\tilde B_k \quad,
\ee
where $G_{\mu \nu}$ transforms like a tensor under Lorentz
transformations. 
With this the equations of motion (\ref{eoms13}, \ref{eoms14}) are
conveniently rewritten as
\be\label{Procas1}
(D^\alpha D_\alpha + m^2)G_{\mu \nu}+ie F_{\alpha \mu}G^\alpha_{\;\nu}
-ieF_{\alpha \nu}G^\alpha_{\;\mu}=0\quad.
\ee
Please note that the tensor field $G_{\mu \nu}$ is a priory
not the field strength of a vector field.
The equation (\ref{Procas1}) is however the quadratic form
of the interacting Proca equations \cite{Proca:1936}
\bea\label{proca1}
D_\mu G^{\mu \nu}&=&-m^2\phi^\nu\quad,\\ \label{proca2}
G_{\mu \nu}&=&D_\mu \phi_\nu-D_\nu \phi_\mu \quad,
\eea
which can be obtained after inserting the first
Proca equation (\ref{proca1}) into the second Proca equation (\ref{proca2}).
Since equation (\ref{proca1}) is purely algebraical for the field $\phi$,
this replacement can be done without loss of generality.
Thus, it has been shown how the model can be matched to the intercating
Proca equations.

\subsection{Arbitrary spin}

For arbitrary spin, the equations of motion 
of the Lagrangian (\ref{Lmaster}) read
\bea\label{arbs1}
D_\mu D^\mu \psi + ig_s e H_{\mu \nu}^s F^{\mu \nu}\psi
+m^2 \psi &=&0 \quad,\\ \label{arbs2}
D_\mu D^\mu \Omega + ig_s e H_{\mu \nu}^{'s} F^{\mu \nu}\Omega
+m^2 \Omega &=&0\quad.
\eea
Where the definition (\ref{Hptensor}) was used.\\

The first equation (\ref{arbs1}) contains 
the $2s+1$ component field $\psi$ living in the $(s,0)$
representation of the Lorentz group.
In \cite{Hurley:1972ju} it was shown that,
this field equation is equivalent
to the relativistic arbitrary spin equation for
fields with $6s+1$ components that live in the
$(s,0)\oplus(s-1/2,1/2)$ representation.\\
The second equation (\ref{arbs2}) contains 
the $2s+1$ component field $\Omega$ living in the $(0,s)$
representation of the Lorentz group.
It also was shown in \cite{Hurley:1972ju}
that equation (\ref{arbs2}) is equivalent
to the relativistic arbitrary spin equation for
fields with $6s+1$ components that live in the
$(0,s)\oplus(1/2,s-1/2)$ representation.
Note that this implies that, for the case of 
spin $3/2$ the equations of motion are expected
to be different from the Rarita-Schwinger equations \cite{Rarita:1941mf} which are
based on a $(1,1/2)\oplus(1/2,1)$ representation of the Lorentz group.\\
The combined equations (\ref{arbs1}, \ref{arbs2})
are equivalent to the parity doubled
equations for arbitrary spin, which in 
the formulation of \cite{Hurley:1972ju}  
contain fields with $12s+2$ components that live in the
$(s,0)\oplus(s-1/2,1/2)\oplus (0,s)\oplus(1/2,s-1/2)$
representation of the Lorentz group.
While the necessity for a parity doubling in \cite{Hurley:1972ju} 
was considered a cumbersome construction \cite{Belinicher:1974am}
it arises naturally for the given Lagrangian
without the necessity of introducing abundant field components 
\footnote{
For higher half integer spin one might further 
generalize the operators
$\Pi^+,\Pi^-$, analogous to the construction in \cite{Brown:1958zz}
by using a $(2s+1)$ dimensional Clifford algebra (\ref{Clifford}).
To investigate on this possibility will be left
to future studies.}.
Advantages of this formulation are
that it allows to work with $2s+1$ field
components only (instead of $\ge6s+1$) and
that it includes the case of spin zero.

Note that the equations of motion for the $\psi$ and $\Omega$
fields may be combined and equivalently
written as single equation for the field $\Psi=(\psi, \Omega)$
\bea\label{nonabars1}
D_\mu D^\mu \Psi + ig_s e \mathcal{H}_{\mu \nu}^s F^{\mu \nu}_a T^a \Psi
+m^2 \Psi &=&0 \quad.
\eea
This equation of motion can also
be derived directly from 
the Lagrangian (\ref{Lmaster2}).

\newpage
\section{Symmetries and conserved quantities}

In this section the potential of the given Lagrangian
will be explored in the context of classical symmetries.
We explicitly discuss:
The energy momentum tensor and probability current,
global and local $U(1)$ - $SU(N)$ symmetries,
and symmetries between fields with different spin.

\subsection{Energy momentum tensor}

Performing a variation
with respect to the 
coordinates, one obtains the energy momentum tensor.
At this point, we are not interested in the dynamics
of the external gauge fields, so the energy momentum tensor is\\
\begin{eqnarray}
{T_{\mu}}^{\sigma} &=& \partial_\mu \bar{\Psi}
                     D^{\sigma}\Psi + D^{\dagger \sigma}\bar{\Psi}
                     \partial_\mu \Psi - \delta^{\sigma}_{\mu}
                     \left( D_{\alpha}^{\dagger}\bar{\Psi}D^{\alpha}\Psi -
                           ig_s e\bar{\Psi}\mathcal{H}_{(s)}^{\alpha\beta}
                           F_{\alpha\beta}^a T_a \Psi - m^2\bar{\Psi}\Psi 
                     \right)\quad.
\end{eqnarray}
We are also interested in the Hamiltonian density 
$T_0^0$ in the non interacting case, in order to perform
below the canonical quatization of fields. 
In the non interacting field theory, the 
energy momentum tensor and the Hamiltonian become
\begin{eqnarray}
{{T_{(n.i.)}}_{\mu}}^{\sigma} &=& \partial_\mu \bar{\Psi}
                     \partial^{\sigma}\Psi + \partial^{\sigma}\bar{\Psi}
                     \partial_\mu \Psi - \delta^{\sigma}_{\mu}
                     \left( \partial_{\alpha}\bar{\Psi}\partial^{\alpha}\Psi
                            - m^2\bar{\Psi}\Psi \right)\quad,\\
\mathcal{H} &=& 2\partial_0 \bar{\Psi}\partial^{0}\Psi - 
               \left(\partial_{\alpha}\bar{\Psi}\partial^{\alpha}\Psi
                            - m^2\bar{\Psi}\Psi \right)\quad.
\end{eqnarray}
%
\subsection{Global and local symmetries}

The Lagrangian (\ref{Lmaster}) is invariant under a global phase
transformation
\be
\psi\rightarrow e^{i\Lambda}\psi,\quad
\Omega \rightarrow e^{i\Lambda}\Omega\quad.
\ee
The corresponding conserved current is
\be\label{conservedC}
J^\mu=-ig\Omega^\dagger(D^\mu \psi)+ig(D^\mu \Omega)^\dagger\psi +c.c. \quad.
\ee
Using the equations of motion (\ref{arbs1}, \ref{arbs2}) and
the identity $\partial_\mu (a\cdot b)=(D_\mu a)b+a D_\mu^* b$
one confirmes that $\partial_\mu J^\mu=0$.
For the case of spin zero the expression (\ref{conservedC})
maintaines its original form.
For the case of spin one half the expression (\ref{conservedC})
can be rewritten by using the equation of motion (\ref{Dirac})
in the more familiar form $J^\mu_{1/2}=i\bar \Psi \gamma^\mu \Psi$.
The current for spin one is 
$J^\mu_1=-2iG^*_{\alpha \beta}D^\mu G^{\alpha \beta}
+2i(D^\mu G_{\alpha \beta})^* G^{\alpha \beta}$, which
by using the equations of motion (\ref{Procas1})
can be shown to fulfill $\partial_\mu J^\mu=0$.

Invariance under
local $U(1)$ gauge transformations is given if the fields transform
like
\be
\psi\rightarrow e^{i e\Lambda(x)}\psi\,,\quad
\Omega\rightarrow e^{i e\Lambda(x)}\Omega\,,\quad
A_\mu \rightarrow A_\mu +\partial_\mu \Lambda(x) \quad.
\ee
This construction can be extended 
to nonabelian $SU(N)$ gauge groups with the generators $T^a$ 
and the gauge coupling $e$
 by writing
\be
\psi\rightarrow e^{ie\Lambda^a(x)T_a}\psi \approx (1+ie \Lambda_a
T^a)\psi\,,\quad
\Omega\rightarrow e^{ie \Lambda^a(x)T_a}\Omega \approx (1+ie \Lambda_a
T^a)\Omega\,,\quad
A^a_\mu \rightarrow A^a_\mu +\partial_\mu \Lambda^a(x) + ie A_{\mu}^b\Lambda^c
f^{a}{ }_{bc} \quad.\label{infTransf}
\ee
%

Note the Lagrangian (\ref{Lmaster2}) allows local and global $SU(N)$
gauge invariance. So we have a conseverd current associated to it.
Indeed, the Lagrangian (\ref{Lmaster2}) is invariant under this transformation
since the spin dependent matrices $H_{(s)}^{\mu\nu}$
act on a different space than the internal $SU(N)$ generators $T_a$.
Computing the conserved current asociated to this symmetry, one finds
\begin{eqnarray}
j_{a}^{\mu} &=& ie\left( (D_{\mu}\Omega)^{\dagger}T^{a}\psi -
\Omega^{\dagger}T^a D_{\mu}\psi 
\right)+ c.c.\quad.
\end{eqnarray}
The gauge transformation in the
formulation with $\Psi$ and $\bar{\Psi}$ is
\begin{eqnarray}
\Psi \rightarrow \left( 
                 \begin{array}{cc} 
                 e^{ie\Lambda^a T_a} & 0 \\ 
                 0 & e^{ig\Lambda^a T_a}
                 \end{array} \right) \Psi \approx
                 \left( 
                 \begin{array}{cc}  
                 \openone + ie\Lambda^a T_a  & 0 \\ 
                 0 & \openone + ie\Lambda^a T_a 
                 \end{array} \right)\Psi \quad,\\
\bar{\Psi} \rightarrow \bar{\Psi}\left( 
                 \begin{array}{cc} 
                 e^{-ie\Lambda^a T_a} & 0 \\ 
                 0 & e^{-ig\Lambda^a T_a}
                 \end{array} \right) \approx
                 \bar{\Psi} \left( 
                 \begin{array}{cc}  
                 \openone - ie\Lambda^a T_a  & 0 \\ 
                 0 & \openone - ie\Lambda^a T_a 
                 \end{array} \right)\quad.
\end{eqnarray}
Thus, for $SU(N)$ the
gauge current may be rewritten as
\begin{eqnarray}
j_{a}^{\mu} &=& ig\left( (D^{\dagger\mu}\bar{\Psi} )
\left(\begin{array}{cc}
 T_{a}&0\\
 0&T_a
\end{array}\right)
\Psi -
                        \bar{\Psi} \left(\begin{array}{cc}
 T_{a}&0\\
 0&T_a
\end{array}\right) D^{\mu}\psi \right)\quad.
\end{eqnarray}

\subsection{Symmetries between fields with different spin}
Since the same Lagrangian is suited for 
any value of the spin $s$,
one is tempted
to believe that it might also provide a useful framework
for symmetries between fields with different spin.

As a proof of concept we wish to write a spin-spin-symmetric Lagrangian for
partners of spin zero and spin one half
by only using the given Lagrangian form. In order to get rid 
of the spin one field $A_\mu$ we switch
off the interactions with external spin one fields by setting $e=0$.
Given the number of degrees of freedom
one can construct a toy model of two spin zero parts 
(fields labeled with ``$_a$'' and ``$_b$'' respectively) and
one spin one half part.
For convenience the spin zero parts of the Lagrangian 
will be written in the
notation (\ref{Lmaster2}), while the spin one half part
will be written in the notation (\ref{Lmaster})
\bea\label{Lsusy}
{\mathcal{L}}&=&{\mathcal{L}}_{0}(\Phi_{a})+{\mathcal{L}}_{0}(\Phi_{b})+{
\mathcal {L}}_{1/2 }(\psi,\Omega)\\ \nonumber
&=&(\partial_\mu \bar\Phi_a \partial^\mu \Phi_a-m^2\bar\Phi_a \Phi_a)
+(\partial_\mu \bar\Phi_b \partial^\mu \Phi_b-m^2\bar\Phi_b \Phi_b)
+(\partial_\mu \Omega^\dagger \partial^\mu \psi-
m^2\Omega^\dagger\psi+c.c)\quad.
\eea
This mix of notations 
has the advantage that both, the scalar fields ($\Phi_{a}$, $\Phi_{b}$)
and the spinor fields ($\Omega$, $\psi$) are two component objects.
Since they all have two components one can write down
transformations that mix the fields:
\bea\label{SusyTransform}
\Phi'_a=&\Phi_a+\alpha_a \psi\;,\quad\,\;\;
\Phi'_b=&\Phi_b+\alpha_b \Omega\,, \\
\nonumber
\Omega'=&\Omega -\alpha_a \gamma^0\Phi_a\,,\quad
\psi'=&\psi-\alpha_b \gamma^0
 \Phi_b\;,
\eea
where $\alpha_a$ and $\alpha_b$ are the infinitesimal real numbered
transformation parameters  and $\gamma^0=\sigma_1$.
Please note that ``$_a$'' and ``$_b$'' are no summation
indices here, they only allow two distinguish different fields.
Even though those transformations mix bosonic and fermionic
fields, they still leave the free Lagrangian (\ref{Lsusy}) invariant,
making (\ref{SusyTransform}) a valid supersymmetry transformation. 
The two conserved currents for the symmetry transformation 
(\ref{SusyTransform})
are
\be\label{SusyCurrents}
j^\mu_a=-\bar \Phi_a \overset{\leftrightarrow}{\partial}^\mu\psi +c.c.\;
,\quad
j^\mu_b=-\bar \Phi_b \overset{\leftrightarrow}{\partial}^\mu\Omega + c.c.\quad.
\ee
By using
the equations of motion one can check that 
the currents (\ref{SusyCurrents}) are conserved
as long as the masses of the fields
are equal.
An unexpected feature of this Lagrangian is that
in contrast to the free Wess- Zumino Lagrangian,
both scalar fields have a kinetic term.
However, we have already seen that $s=0$ contains
a ghost fields whose kinetic term could
be canceled by imposing an additional constraint.
If one does this, the difference with respect to 
the Wess- Zumino Lagrangian disappears.

The formalism (\ref{Lsusy},\,\ref{SusyTransform}) relates the $s=0$ fields
$\Phi_a$ and $\Phi_b$ to the $s'=1/2$ fields $\psi$, $\Omega$. 
Since $H^{(0)}=0$ it can not be generalized in a straight forward way to
the interacting case.
The same formalism
also works for fields  $\Phi_a$ and $\Phi_b$ with
arbitrary spin $s$  in the $2(2s+1)$ dimensional representation. The
corresponding
partner fields $\psi$, $\Omega$ in the $2s'+1$ dimensional representation have
the
spin
\be
s'=2s+\frac{1}{2}\quad.
\ee
One sees that $s'$ is always half integer valued,
independent of the spin value of $s$.
Thus while the above example $(s=0,\,s'=1/2)$ looks similar to
supersymmetry, the following spin pair
$(s=1/2,\,s'=3/2)$ involves only fields of half integer spin.
A possible symmetry between fermionic fields of different
spin is a feature which this formulation shares with the
much more general formulation \cite{Buchbinder:2007ak,Buchbinder:2009pa}.

Although this symmetry has nice features in the free particle case,
a straight forward generalization to the interacting case 
with a fixed external field $A_\mu$ seems to be
doubtful.
For a given representation of $s>0$ with ${\mathcal{H}}_{\mu \nu}^{s}$ 
of the type (\ref{mathcalH}), 
imposing a cancellation with the corresponding
terms $\sim H_{\mu \nu}^{2s+1/2}$ in the representation
(\ref{originalH}) leads to the condition
\be
H_{\mu \nu}'^{s}=-H_{\mu \nu}^{s}\quad.
\ee
Thus, by virtue of the definitions (\ref{Htensor}, \ref{Hptensor}), this
symmetry will always be broken in
the presence of an external magnetic field 
$B_k=1/2\epsilon_{ijk} F_{ij}$. Whether this breaking
can be cured by a simultaneous transformation of the external
field remains to be seen.
\newpage
\section{A Quantum Field Theroy}
It will be shown that the model can be used
to exemplify various fundamental topics of 
quantum field theory like:
 Field quantization in connection with
    the spin statistics theorem and the
 derivation of Feynman rules.

\subsection{Quantization}
\label{spinstat}
For convenience the quantization of the free fields will be
carried out with the Lagrangian (\ref{Lmaster2}).
The canonical momenta of this Lagrangian are
\be
\Pi=\frac{\partial {\mathcal{L}}}{\partial \dot\Psi}=\dot{\bar\Psi};
\quad
\bar\Pi=\frac{\partial {\mathcal{L}}}{\partial \dot{\bar\Psi}}=\dot{\Psi}
\ee
and their quantization dictates the following (anti-)commutation relations.
At this point one can not know whether commutation or anti-commutation applies, so 
one has to leave open both possibilities by assigning a $-$ or a $+$ to the bracket
\bea\label{psicommut}
\left[\Psi(x),\Pi(x') \right]_{\pm}&=&i\delta^3(x-x')\quad,\\
\nonumber
\left[\bar\Pi(x),\bar\Psi(x') \right]_{\pm}&=&-i\delta^3(x-x')\quad.
\eea
The fields can be expanded in terms of Fourier components
$f_k(x)=e^{-ikx}/\sqrt{(2\pi)^32\omega_k}$,
momentum space field operators $a_\sigma(k),\;b_\sigma(k)$, and normalized
momentum space solutions $u_\sigma(k),\,\nu_\sigma(k)$:
\bea\label{ExpandPsi}
\Psi(x)&=&\int d^3k \frac{1}{\sqrt{(2\pi)^32\omega_k}}
\sum_\sigma 
(u_\sigma(k) f_k(x)a_{\sigma}(k)+
\nu_\sigma(k) f_k^*(x)b^\dagger_\sigma(k))\quad,\\ \nonumber
\bar\Psi(x)&=&\int d^3k \frac{1}{\sqrt{(2\pi)^32\omega_k}}
\sum_\sigma 
(\bar u_\sigma(k) f^*_k(x)a^\dagger_{\sigma}(k)+
\bar\nu_\sigma(k) f_k(x)b_\sigma(k))\quad.
\eea
The momentum space free field solutions $u_\sigma,\,\nu_\sigma$ are
normalized to the absolute value of one, but at this
point it will be left open which sign this normalization is supposed
to carry. This sign ambiguity
is parameterized by introducing
$n_u$ and $n_\nu$ which can be either zero or one
\be\label{Normu}
\bar u_\sigma(k) u_{\sigma'}(k)=(-1)^{n_u}\delta_{\sigma \sigma'}\;,\quad
\bar \nu_\sigma(k) \nu_{\sigma'}(k)=(-1)^{n_\nu}\delta_{\sigma \sigma'}\quad.
\ee
In order to find the (anti)-commutation relations
for the field operators, the Fourier expansion (\ref{ExpandPsi})
has to be inverted. By using the orthonormality relations
\be\label{fortho}
\int d^3x
f_k^*(i\overset{\leftrightarrow}{\partial}_0)f_{k'}=\delta^3(k-k'),\quad
\int d^3x
f_k(i\overset{\leftrightarrow}{\partial}_0)f^*_{k'}=-\delta^3(k-k')\quad,
\ee
one finds
\bea\label{avonPsi}
a_\sigma(k)&=& (-1)^{n_u}
\int d^3x \sqrt{(2\pi)^3 2 \omega_k}
\bar u_\sigma f^*_k(x)(i\overset{\leftrightarrow}{\partial}_0)\Psi(x)\quad,\\
\nonumber
a^\dagger_\sigma(k)&=& (-1)^{n_u}
\int d^3x \sqrt{(2\pi)^3 2 \omega_k}
\bar
\bar\Psi(x)(i\overset{\leftrightarrow}{\partial}_0)f_k(x)u_\sigma(k)\quad,\\
\nonumber
b_\sigma(k)&=& -(-1)^{n_\nu}
\int d^3x \sqrt{(2\pi)^3 2 \omega_k}
\bar \bar\Psi(x)(i\overset{\leftrightarrow}{\partial}_0)f^*_k(x)\nu_\sigma(k)
\quad,\\
\nonumber
b^\dagger_\sigma(k)&=& -(-1)^{n_\nu}
\int d^3x \sqrt{(2\pi)^3 2 \omega_k}
\bar \nu_\sigma f_k(x)(i\overset{\leftrightarrow}{\partial}_0)\Psi(x)\quad.
\eea
Using (\ref{psicommut}, \ref{fortho}, and \ref{avonPsi})
one deduces the (anti)-commutation relations for the momentum space
field operators
\bea\label{ComRel}
\left[a(p,\sigma),a^\dagger(p', \sigma') \right]_\pm&=&
(-1)^{n_u}(2\pi)^3 2 k_0\delta^3({\bf{p}-\bf{p'}}) \delta_{\sigma\sigma'}
\quad,\\
\nonumber
\left[b^\dagger(p,\sigma),b(p', \sigma') \right]_\pm&=&
-(-1)^{n_\nu}(2\pi)^3 2 k_0\delta^3({\bf{p}-\bf{p'}})
\delta_{\sigma\sigma'}\quad .
\eea
Now the free field Hamiltonian 
will be derived.
The free field Hamiltonian density for (\ref{Lmaster2}) reads
\be\label{Hdens}
{\mathcal{H}}
=\Pi \dot\Psi+\dot{\bar\Psi} \bar\Pi-{\mathcal{L}}=
(\partial_0 \bar\Psi)
(\partial_0 \Psi)+
(\partial_i \bar\Psi)
(\partial_i \Psi)+
m^2 \bar\Psi \Psi \quad.
\ee
In terms of the momentum space operators this Hamiltonian
is
\be\label{QHamilton}
H=\int d^3x \;{\mathcal{H}}=
\int d^3k \frac{k_0^2+k_i^2+m^2}{4(2\pi)^3 \omega_k^2}
((-1)^{n_u}a_\sigma^\dagger(k) a_\sigma(k)+
(-1)^{n_\nu}b_\sigma(k)b_\sigma^\dagger(k))\quad.
\ee
In order to have a finite normal ordered expression
one has to use $b b^\dagger=\mp b^\dagger b \pm [b^\dagger,b]_{\pm}$,
dropping the infinite contribution from the (anti)-commutator
leads to
\be\label{QHamilton2}
:H:=\int d^3k \frac{k_0^2+k_i^2+m^2}{4(2\pi)^3 \omega_k^2}
((-1)^{n_u}a_\sigma^\dagger(k) a_\sigma(k)
\mp(-1)^{n_\nu}b^\dagger_\sigma(k)b_\sigma)\quad.
\ee
Imposing a positivity condition on the normal ordered Hamiltonian
(\ref{QHamilton2}) allows to determine the
sign of the wave function normalization in terms of the statistics
\be\label{NormStat}
n_u=0\quad
{\mbox{and}}\quad
n_\nu=\left\{
\begin{array}{cc}
1& \mbox{for Fermi statistics}\\
0& \mbox{for Bose statistics}
\end{array}
\right.
\quad.
\ee
Thus, one has on the one hand 
for any statistics $\bar u_\sigma(k) u_{\sigma'}(k)=\delta_{\sigma \sigma'}$
and on the other hand
$\bar \nu_\sigma(k) \nu_{\sigma'}(k)=\delta_{\sigma \sigma'}$ 
for boson-statistics and
$\bar \nu_\sigma(k) \nu_{\sigma'}(k)=-\delta_{\sigma \sigma'}$ 
for fermion-statistics.
Note that this result also allows to write the
(anti)-commutation relations for the momentum space
operators (\ref{ComRel}) in their familiar and form
which turns out to be independent of the normalizations $n_\nu$ and $n_u$ 
\bea\label{ComRel2}
\left[a(p,\sigma),a^\dagger(p', \sigma') \right]_\pm&=&
(2\pi)^3 2 k_0\delta^3({\bf{p}-\bf{p'}}) \delta_{\sigma\sigma'}
\quad,\\
\nonumber
\left[b(p,\sigma),b^\dagger(p', \sigma') \right]_\pm&=&
(2\pi)^3 2 k_0\delta^3({\bf{p}-\bf{p'}})
\delta_{\sigma\sigma'}\quad .
\eea

\subsection{Statistics}
\label{spinstat2}
Given the connection between normalization
and statistics (\ref{NormStat}), the key for
finding the relation between spin and statistics in this model
lies in finding the relation between spin and normalization:
The possible values of the normalizations
$\bar u_\sigma u_{\sigma'}$ and $\bar \nu_\sigma \nu_{\sigma'}$
are determined by the $2(2s+1)$ eigenvalues of the matrix $\gamma_0$.
By diagonalizing this matrix as it is defined in (\ref{gamma0})
one finds that it has $2s+1$ eigenvalues $e_{u_i}=1$ and 
$2s+1$ eigenvalues $e_{\nu_i}=-1$.
A proof of this normalization is given in the appendix \ref{appNorm}.
According to the definition of the normalization (\ref{Normu})
those eigenvalues dictate $n_u=0$ and $n_\nu=1$
independent  of the spin $s$.
Thus, finally due to (\ref{NormStat}) one has no
other choice than to conclude that all fields
of the Lagrangian (\ref{Lmaster}) obey anti-commuting
Fermi statistics, no matter which spin $s$ they carry.
For the sake of completeness we perform a number of
consistency checks on this result:
\begin{itemize}
\item The charge of a Dirac field is proportional to $\Psi^\dagger \Psi$
 which can be checked for the arbitrary spin fields at the level of quantization
\be
Q_{D}=\int d^3x :\Psi^\dagger(x) \Psi(x):
=\int d^3k \frac{1}{32 \pi^3}\frac{1}{\omega_k m}\sum_\sigma
(a^\dagger_\sigma(k)a_\sigma(k)\mp b^\dagger_\sigma(k) b_\sigma(k))\quad,
\ee
where $u^\dagger_\sigma(k)u_{\sigma'}(k)=\delta_{\sigma \sigma'}\omega_k/m$
and $\nu^\dagger_\sigma(k)\nu_{\sigma'}(k)=\delta_{\sigma \sigma'}\omega_k/m$
has been used and where the upper sign refers to anti-commutation and the lower
sign refers to commutation relations. Asking for the existence of positive and
negative charges one observes that only anti-commutation relations can
provide a physically acceptable result.
Thus, also the possible existence of electrical
charge for this model dictates anti-commutation
relations.
\item An important result from the previous section
is the conserved current (\ref{conservedC})
that is following from the $U(1)$ symmetry
of the Lagrangian. This current gives rise to a conserved 
charge
\be
Q_{N}=:\int d^3x \bar \Psi i \overset{\leftrightarrow}{\partial^0} \Psi:\quad.
\ee
Using the (anti)-commuting quantization rules one obtains
\be
Q_{N}=\int d^3x \frac{1}{16 \pi^3}\frac{1}{\omega_k}\sum_\sigma
(a^\dagger_\sigma(k)a_\sigma(k)\mp b^\dagger_\sigma(k) b_\sigma(k))\quad.
\ee
The only way to allow for negative charges is by picking the upper 
sign, which corresponds to anti-commuting field operators.
Thus, also the possible existence of electrical
charge for this model dictates anti-commutation
relations.
\item
 In order to check whether
the creation and annihilation operators have been assigned
correctly one probes that
\be\label{creationcheck}
[x,:H:]=\omega x
\left\{
\begin{array}{cc}
 \omega > 0 & \mbox{for $x$ being annihilation operator}\\
 \omega < 0 & \mbox{for $x$ being creation operator}
\end{array}\right.\quad.
\ee
Performing this check for example for $b^\dagger_\sigma$
one finds
\be
[b^\dagger_\sigma(k),:H:]=\pm (-1)^{n_\nu}\frac{1}{2}
(k_0^2+k_i^2+m^2)b^\dagger_\sigma(k)\quad.
\ee
 Comparing this with (\ref{creationcheck}) one
 finds that the result (\ref{NormStat}) is confirmed.
This means that the creation and 
annihilation operators are assigned correctly.

 \item 
The ingredient of the usual spin-statistics theorem
that remains to be checked is causality.
Causality can be studied by revising
whether the expression
$\bra{0} \left[ \Psi_a(x),\bar\Psi_b(y) \right]_{\pm}\ket{0}$
vanishes for spacelike separations. Using 
the relations (\ref{ComRel2}) one finds
\be\label{causi}
\bra{0} \left[ \Psi_a(x),\bar\Psi_b(y) \right]_{\pm} \ket{0} =
\int d^3k \sum_\sigma \frac{1}{(2\pi)^3 2 \omega_k}
\left(
u_{\sigma,a}(k)\bar u_{\sigma,b}(k)e^{-ik(x-y)}
\pm
\bar \nu_{\sigma,b}(k) \nu_{\sigma,a}(k)e^{ik(x-y)}
\right)\;.
\ee
For representations that fulfill the Clifford algebra (\ref{Clifford}) 
one can simplify this expression by using (\ref{cmpltu} and \ref{cmpltv})
\be
\bra{0} \left[ \Psi_a(x),\bar\Psi_b(y) \right]_{\pm}\ket{0}=
{\mathcal{A}}_{ab}(\partial)
\int d^3k  \frac{1}{(2\pi)^3 2 \omega_k}
\left(
e^{-ik(x-y)}\mp e^{ik(x-y)}
\right)\quad,
\ee
where ${\mathcal{A}}_{ab}(\partial)$ is a matrix valued derivative
operator. One sees that for
space-like separations $(x-y)^2<0$ this expression only
vanishes for the upper sign and is non-zero for the lower sign.
For a more general case that also includes
(\ref{cmpltus1}, \ref{cmpltvs1})
the situation is more complicated and we restrict
to the calculation of the trace of the expression (\ref{causi}).
By using  $\sum_{\sigma, a} \bar u_{\sigma,a} u_{\sigma, a}=2s+1$
and $\sum_{\sigma, a} \bar \nu_{\sigma,a} \nu_{\sigma, a}=-2s-1$
one finds
\be
\sum_a \bra{0} \left[ \Psi_a(x),\bar\Psi_a(y) \right]_{\pm}\ket{0}=
\int d^3k  \frac{2s+1}{(2\pi)^3 2 \omega_k}
\left(
e^{-ik(x-y)}\mp e^{ik(x-y)}
\right)\quad.
\ee
Again, one sees that for
space-like separations this expression only
vanishes for the upper sign and is non-zero for the lower sign.
Thus, also the causality condition dictates anti-commutation relations 
for any spin.
\end{itemize}
Thus, performing a careful quantization and normalization procedure
it has been shown that a Lagrangian of the type (\ref{Lmaster})
can only contain fields that obey anti-commutation relations,
independent of the actual spin of those fields.
Since this is in contradiction to the 
spin-statistics theorem for physical fields,
it implies that the spin $0,1,\dots$ fields in this
theory are necessarily ghost fields.
However, one would tend to call
them ``good'' ghosts since they do not give rise to a negative
Hamiltonian density. 
Whether those ghost fields with
spin greater than zero proof to be as useful as for example the spin zero 
Fadeev-Popov ghosts remains to be seen.

\subsection{Feynman rules and Issues}
\begin{figure}[*h!t!p!]
\begin{minipage}{9cm}
\includegraphics[width=1\textwidth]{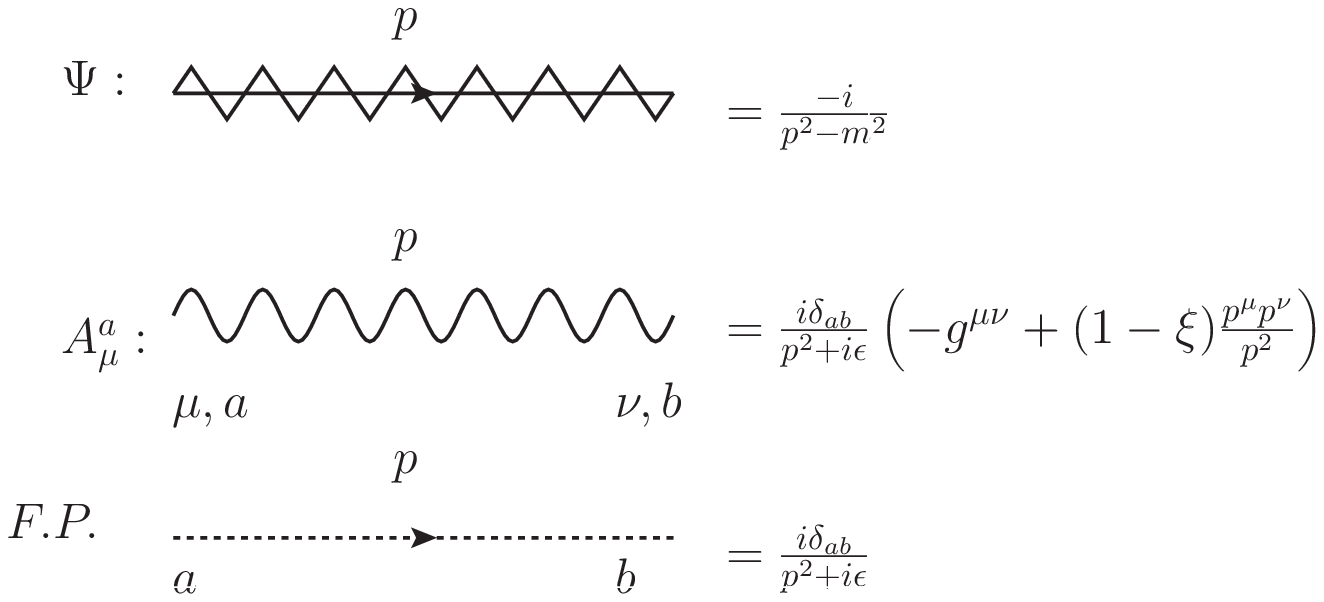}
\caption{Propagators for the Full Lagrangian}
\label{prop}
\end{minipage}
\begin{minipage}{8cm}
\includegraphics[width=1\textwidth]{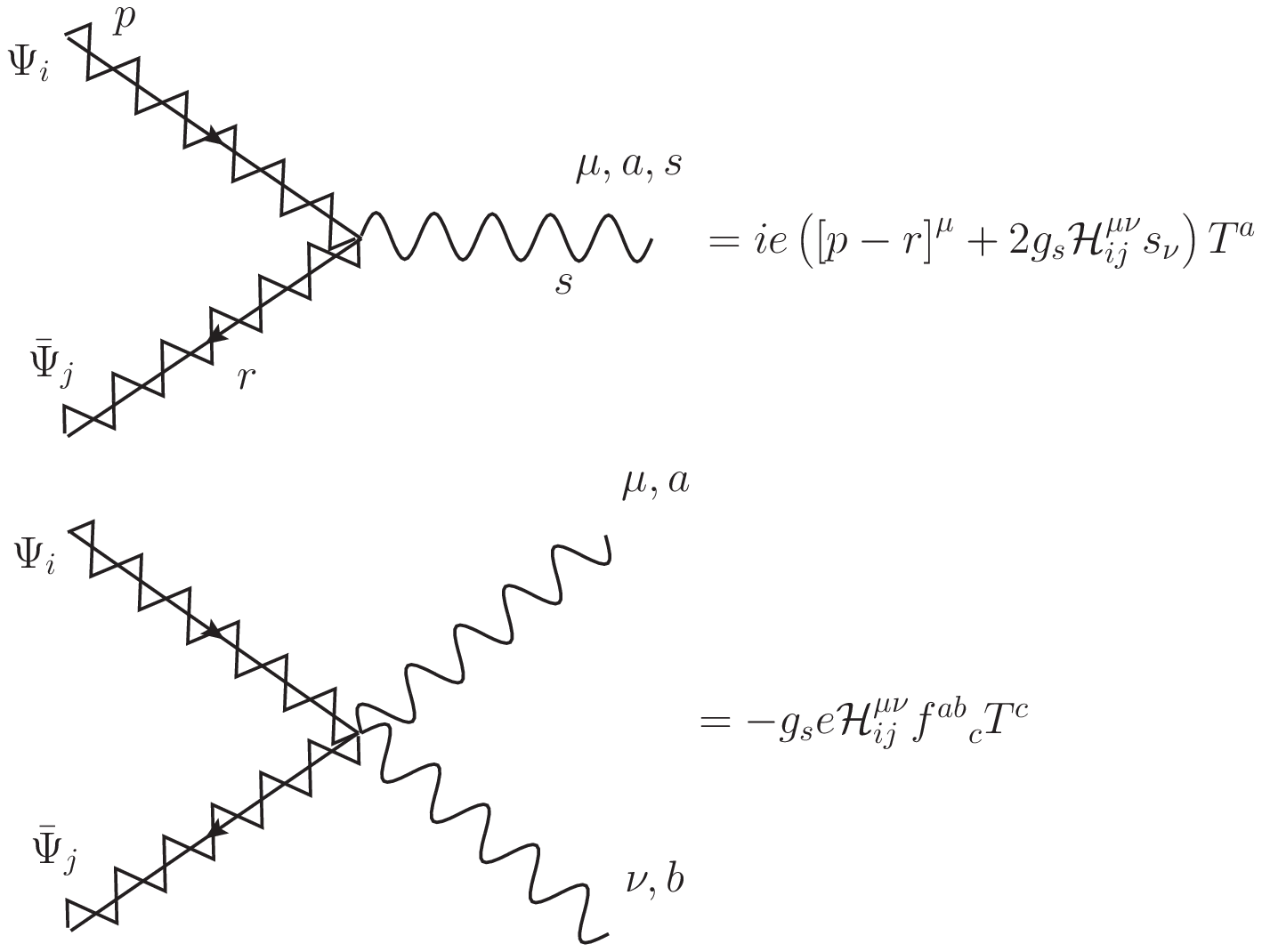}
\caption{Interactions for $\Psi$, $\bar{\Psi}$ fields. All the particles 
left to the interaction vertex are incoming and all the particles right to the 
interaction vertex are outgoing.}
\label{vertex}
\end{minipage}
\end{figure}

\begin{figure}[*h!t!p!]
\includegraphics[width=0.6\textwidth]{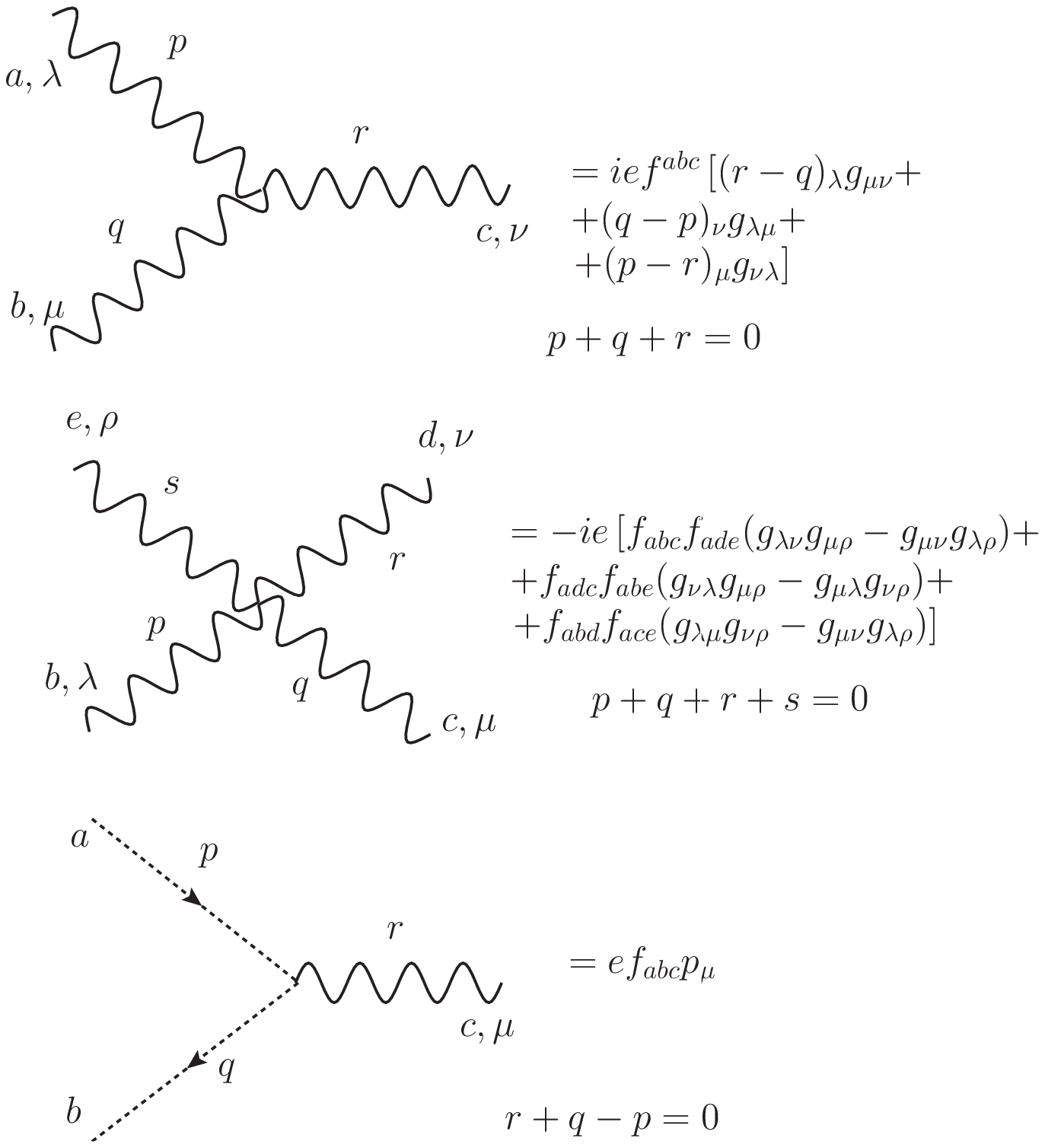}
\caption{Interactions for $A_{\mu}^a$ and $\eta$ fields. Here 
the convention is that all 
the particles are incoming.}
\label{secint}
\end{figure}

In order to derive the Feynman rules for the theory
and to be consistent with a theory
with full particle content one has to add
a dynamic term for the gauge field, and
a gauge fixing term plus Fadeev Popov ghost terms.
Nevertheless, we are interested in the
interactions of $\Psi$ and $\bar{\Psi}$,
and gauge fixing and ghosts are not
interacting directly with our fields, so we
put their Feynman rules just for
sake of completeness.
When computing the Feynman rules
directly from a path integral formulation one
has to add Grassmann numbers as 
sources for $\Psi$ and $\bar{\Psi}$ in 
the effective action. Remember that
$\Psi$ and $\bar{\Psi}$ obey anti-commutative
relations only. Thus, the generating functional 
may be written as
\begin{eqnarray}
\mathcal{Z}\left[\bar{\lambda},\lambda,\dots \right] &=& \int \mathcal{D}\bar{\Psi}\mathcal{D}\Psi
                \bar{\mathcal{D}}A\mathcal{D}\eta^{*}\mathcal{D}\eta 
                \exp \left(i\int d^4 x \left\{ \mathcal{L}_{n.a.} - 
                \frac{1}{4}F_{\mu\nu}^a F^{\mu\nu}_a -
                \frac{1}{2\xi}(\partial_{\mu}A^{\mu}_a)^2 - 
                \partial_{\mu}\eta^{*}_{a} (\partial^{\mu}\eta^a + 
                e f^{a}_{bc}\eta^b A^{\mu c}) +  \right. \right. \nonumber \\
            &+& \left. \left. \bar{\lambda}\Psi + \bar{\Psi}\lambda 
+ S \right\} \right) \quad,
\end{eqnarray}

where $\mathcal{L}_{n.a.}$ is the Lagrangian (\ref{Lmaster2}) 
and $S$
are the sources for gauge and Fadeev-Popov 
ghosts fields $\eta$. 
The dots in $\mathcal{Z}\left[\bar{\lambda},\lambda,\dots\right]$
label the other sources included in S.
The generating functional may be also written as
\begin{eqnarray}
\mathcal{Z}\left[\bar{\lambda},\lambda,\dots\right] &=& \exp\left( i\int \mathcal{L}_I
               \left( \frac{\delta}{\delta \bar{\lambda}},\frac{\delta}{\delta \lambda}, \dots \right) \right)Z_{0} \\
Z_{0} &=& \int \mathcal{D}\bar{\Psi}\mathcal{D}\Psi
                \bar{\mathcal{D}}A\mathcal{D}\eta^{*}\mathcal{D}\eta 
                \exp \left(i\int d^4 x \left\{ \mathcal{L}_0 (\bar{\Psi},\Psi) - 
                \frac{1}{4}(\partial_\mu A_{\nu}^a -  \partial_\mu A_{\nu}^a)(\partial^\mu A^{\nu}_a -  \partial^\mu A^{\nu}_a) \right. \right. \nonumber \\
      &-&       \frac{1}{2\xi}(\partial_{\mu}A^{\mu}_a)^2 - 
                \partial_{\mu}\eta^{*}_{a}\partial^{\mu}\eta^a + 
                \left. \left. \bar{\lambda}\Psi + \bar{\Psi}\lambda 
+ S \right\} \right)\quad,
\end{eqnarray}
where $\mathcal{L}_0$ is the
non interacting Lagrangian for $\Psi$ and 
$\bar{\Psi}$. All the interactions are kept
in the $\mathcal{L}_I$ term
\begin{eqnarray}
\mathcal{L}_I &=& ie\bar{\Psi}A_{\mu}^a T_a \partial_{\mu}\Psi - 
                  ie\partial_{\mu}\bar{\Psi}A_{\mu}^a T_a \Psi - 
                  ig_s e\bar{\Psi}\mathcal{H}^{\mu\nu}_{(s)}F_{\mu\nu}^a T_a \Psi -
                  \frac{1}{4}\left< AAA + AAAA \right> 
- \left< \eta \eta A \right>\quad,
\end{eqnarray}
where the quantities in brackets
are cubic and quartic 
interactions of $A_{\mu}^a$ fields and the
Fadeeev-Popov ghost gauge field vertex, as
used in a conventional Yang Mills action.
The Feynman rules are derived from
\begin{eqnarray}
\mathcal{W}\left[ \bar{\lambda},\lambda,\dots\right] &=& -i\ln \left(
\mathcal{Z}\left[ \bar{\lambda},\lambda,\dots \right] \right)\quad,
\end{eqnarray}

whose functional derivatives with respect to the sources 
give rise to the connected Green functions.
Once the Lagrangian is split into
non interacting and interacting parts,
non interactions will give rise propagators
and interacting parts will give rise to vertices.
Propagators are shown in figure \ref{prop}
and vertices are shown in figure \ref{vertex}.
Note that, for particles with integer spin, propagators
serve just as an internal line. This
is a consequence of anomalous relation between
spin and statistics of the $\Psi$ fields for
integer spin. Half integer fields may,
however, appear as internal and external lines.
Another interesting feature of this formulation
is that the propagators of $\Psi$, and $\bar{\Psi}$
are of the Klein Gordon type which means that
they do not carry any spin dependence.
Still, the spin dependence in this formulation
appears in the interaction term $\sim \mathcal{H}^{\mu\nu}$.
For the case of spin one half it remains to be investigated in a beyond tree level calculation,
whether this leads to some measurable difference with respect to the
Dirac action. 
A further peculiarity of those Feynman rules is that
the quartic interaction is antisymmetric
under the interchange of $\mu$ and $\nu$ indices, and 
it is also antisymmetric 
under the interchange of $a$ and $b$ indices, so permuting gauge fields
does not change the sign of the vertex.
The interchange of fields $\bar{\Psi}$
and $\Psi$ does not affect the global sign of
the coupling. Note that this applies to the
cubic interaction too. Feynman rules of 
interactions of $\Psi$ and $\bar{\Psi}$
have spin dependence only 
given in the $\mathcal{H}^{\mu\nu}$ matrix.
For sake of completeness, Feynman rules
for gauge and Fadeev-Popov ghosts are
derived and they are shown in figure \ref{secint}.
The simplicity of the toy Lagrangian (\ref{Lmaster})
has its price. 
\begin{itemize}
 \item
As it has been seen from the quantization
section (\ref{spinstat2}) the bosonic fields 
are obliged to be ghost fields due to their
anti-commuting statistics.
\item
One would have to perform further checks of the
type of \cite{Buchdahl:1958xv,Velo:1970ur,Buchdahl:1982ni}
in order to see whether the coupling to external
gauge fields leads to inconsistencies.
Even if there are no further problems with
the coupling to standard gauge fields, the
coupling of this model to gravity will most
certainly lead to the type of
inconsistencies described in \cite{Deser:2001dt}
\footnote{We thank Stanley Deser for pointing this out.}.
\item
If one restricts to equations
involving field strengths or ``minimal fields'' 
the much more general formulation
of \cite{Siegel:1986de,Siegel:1986zi,Siegel:1988yz}
also takes an extremely simple form
(after eliminating the auxiliary fields). 
However, when going beyond tree level,
such reduced formulations (like the
one presented here), are expected to
lead to further problems with ghost fields
at the loop level
\footnote{We thank Warren Siegel for pointing this out.}.
\end{itemize}

\clearpage
\section{Conclusions}
We report on an extended student project where
a generic Lagrangian (\ref{Lmaster}) and 
a generic Lorentz transformation (\ref{Ltrans}) is investigated
(similar to previous studies \cite{Koch:2011hc}).
This Lagrangian allows to obtain the Klein-Gordon equation,
the Dirac equation, and the Proca equations
as special cases.
The choice of which theory one wants to get
is done by fixing a linear operator (\ref{Htensor}) to a special
representation of SU(2) and by
identifying the fields $\Omega$ and $\psi$
according to this choice.
It is found that for arbitrary spin $s>0$,
the resulting equations of motion are identical to
the equations derived in \cite{Hurley:1972ju}.
The main advantage of the Lagrangian (\ref{Lmaster})
is its simple form which allows 
to introduce and to study simultaneously field equations 
and fundamental quantum field theoretical concepts for various spin 
without overly heavy or abstract mathematical constructions.

In the discussion, various aspects of
the model are explored:
Local and global symmetries of the Lagrangian
are discussed and the conserved currents are calculated.
After this, the model is used in order
to construct a simple Lagrangian
which is invariant under transformations between
spin zero fields $\Psi_a,\, \Psi_v$ and spin one half partner fields ($\psi$,
$\Omega$).
It is also shown that this symmetry between free fields of different spin 
can be generalized to a symmetry between 
spin $s$ fields $\Psi_a,\, \Psi_b$ and spin $2s+1/2$
partner fields $\psi$, $\Omega$.

Then, quantization in the free field case is reviewed.
The discussion reveals that all fields,
that are described by this Lagrangian have to
obey Fermi statistics
\bea
\left[a(p,\sigma),a^\dagger(p', \sigma') \right]_+&=&
(2\pi)^3 2 k_0\delta^3({\bf{p}-\bf{p'}}) \delta_{\sigma\sigma'}
\quad,\\
\nonumber
\left[b(p,\sigma),b^\dagger(p', \sigma') \right]_+&=&
(2\pi)^3 2 k_0\delta^3({\bf{p}-\bf{p'}})
\delta_{\sigma\sigma'}\quad .
\eea
This holds of course also 
for the spin zero and spin one fields discussed before.
Thus, even though the equations of motion of
this Lagrangian for the case of spin zero and spin one
are the familiar ones, the quantized Lagrangian is fundamentally
different.
After performing a number of consistency checks we
conclude, by the virtue of the spin-statistics
theorem, that in this Lagrangian the fields with spin $0,1,2,\dots$
are necessarily ghost fields. 
On the other hand, this proof also shows that physical fermionic
fields are well described within this model.
Finally, based on this consistent quantization, the Feynman rules 
for the matter fields with arbitrary spin are written down
and further issues are pointed out
\footnote{
The detailed  formulation of the
cases $s=3/2$ and $s=2$ will be left for future studies.}.\\

Many thanks to M.A. Diaz, M. Ba\~nados, C. Valenzuela, 
and the Atlas-Andino group for valuable 
hints and discussions. 
The work of B. K. was supported by CONICYT
project PBCTNRO PSD-73 and FONDECYT project 1120360. The work of
N. R. was supported by CONICYT scholarship.
\newpage
\section{Appendix}
\subsection{Proof of the field normalization}
\label{appNorm}
For all $j,k: -s \dots s$,
we seek to solve a system of equations that contains
\bea\label{condi1}
 \bar\nu_j \nu_k&=&\pm \delta_{jk},\\
 \bar u_j u_k &=&\delta_{jk}\\
u^\dagger_j \nu_k&=&0\\
\nu^\dagger_j u_k&=&0\\
\bar u_j \nu_k&=&0\\
\bar\nu_j u_k&=&0\quad.
\eea
The positive
quantities $u^\dagger_j u_k$ and $\nu^\dagger_j \nu_k$
are possibly dependent on their Lorentz frame.
We assume that they agree in the rest frame
with the absolute value of the corresponding Lorentz invariant quantities
\bea
u^\dagger_j u_k|_{rest}&=&\delta_{jk}\\ \label{condi8}
\nu^\dagger_j \nu_k |_{rest}&=&\delta_{jk}\quad.
\eea

The $8(2s+1)^2$ equations (\ref{condi1}-\ref{condi8})
allow to determine the complex ``vectors''
$u_j=(a_{uj};b_{uj})$ and $\nu=(a_{\nu j}; b_{\nu j})$,
where the $a_{uj}, b_{uj}, a_{\nu j}, b_{\nu j}$ have $2s+1$ complex
components and $u_j, \nu_j$ have $2(2s+1)$ complex components.

Without loss of generality one can make the choice
\bea
\nu_1=(a_{\nu1},0, \dots ; b_{\nu1},0,\dots),&
\nu_2=(0,a_{\nu2}, \dots ; 0,b_{\nu2},\dots),&\dots \\
u_1=(a_{u1},0, \dots ; b_{u1},0,\dots),&
u_2=(0,a_{u2}, \dots ; 0,b_{u2},\dots),&\dots
\eea
one observes that now most of the $8(2s+1)^2$ equations
 (\ref{condi1}-\ref{condi8}) are solved, leaving us with
only eight diagonal equations $j=k$ for revery single spin state $j$.
\bea\label{condi1b}
 \bar\nu_j \nu_j&=&\pm 1,\\
 \bar u_j u_j &=&1\\
u^\dagger_j \nu_j&=&0\\
\nu^\dagger_j u_j&=&0\\
\bar u_j \nu_j&=&0\\
\bar\nu_j u_j&=&0\\
u^\dagger_j u_j|_{rest}&=&1\\ \label{condi8b}
\nu^\dagger_j \nu_j |_{rest}&=&1\quad.
\eea
This system 
has {\bf {no}} solution for $\bar\nu_j \nu_j=+1$.
In contrast it is solvable for $\bar\nu_j \nu_j=-1$
giving
\bea
\nu_1=\frac{1}{\sqrt{2}}(-1,0,\dots;1,0,\dots),&
\nu_2=\frac{1}{\sqrt{2}}(0,-1,\dots;0,1,\dots),&\dots \\
u_1=\frac{1}{\sqrt{2}}(1,0,\dots;1,0,\dots), &
u_2=\frac{1}{\sqrt{2}}(0,1,\dots;0,1,\dots),&\dots
\eea
The result holds in the rest frame. Since a Lorentz
boost can not change the sign of a normalization
and since $\bar \nu_j \nu_j$ is Lorentz invariant,
it holds in any Lorentz frame.
This proves that $\bar\nu_j \nu_k=-\delta_{jk}$ is 
the only possible field normalization.

\subsection{Completeness relations for $u$ and $\nu$}
As one can note also in the section \ref{spinstat}
these spinors are normalized respect to the
adjoint spinors $\bar{u}$ and
$\bar{v}$ in any reference frame. Nevertheless,
they are normalized with the respective
hermitian transpose just in the restframe.
This is because we want to measure a positive
definite probability density in the particle
restframe. Combining these two facts, one
has in the particle restframe:
\begin{eqnarray}
\bar{u}_\sigma u_\sigma &=& 1 \label{unorm2}\\
u^{\dagger}_\sigma (0) u_\sigma (0) &=& 1 = \bar{u}_\sigma u_\sigma \\
\bar{\nu}_\sigma \nu_\sigma &=& -1 \label{vnorm2}\\
\nu^{\dagger}_\sigma (0) \nu_\sigma (0) &=& 1 = \bar{\nu}_\sigma \nu_\sigma 
\end{eqnarray}

Note that the sum convention is used
in equations (\ref{unorm2}) and (\ref{vnorm2}), for
spinor index $\sigma$.
Using the definition of the adjoint spinors
$\bar{u}_\sigma = u^\dagger_\sigma \gamma_0$ and 
$\bar{v}_\sigma = v^\dagger_\sigma \gamma_0$, one can split
above equations
\begin{eqnarray}
u^{\dagger}_\sigma (0) (\gamma_0 - \openone) u_\sigma (0) &=& 0\\
\nu^{\dagger}_\sigma (0) (\gamma_0 + \openone) \nu_\sigma (0) &=& 0\quad.
\end{eqnarray}

Now, the spinor basis for $u$ and $v$ is chosen
such they are eigenvectors of $\gamma_0$
with $\pm1$ eigenvalues respectively. This
implies that
\begin{eqnarray}
(\gamma_0 - \openone)u_\sigma &=& 0\\
u_\sigma^{\dagger}(\gamma_0 - \openone) &=& \bar{u}_\sigma (\openone - \gamma_0) = 0 \Rightarrow \nonumber \\
u_\sigma \bar{u}_\sigma \gamma_0 &=& u_\sigma \bar{u}_\sigma \\
\gamma_0 u_\sigma \bar{u}_\sigma  &=& u_\sigma \bar{u}_\sigma \Rightarrow \\
u_\sigma \bar{u}_\sigma \gamma_0 &=& \gamma_0 u_\sigma \bar{u}_\sigma = u_\sigma \bar{u}_\sigma
\end{eqnarray}

That means in the particle restframe, the 
projector $u_\sigma \bar{u}_\sigma$, commute with
$\gamma_0$, in other words it can be only
a linear combination of $\gamma_0$ and
the identity into this frame. So, one gets the ansatz
\begin{eqnarray}
u_\sigma \bar{u}_\sigma &=& A \gamma_0 + B \openone \nonumber \\
u_\sigma \bar{u}_\sigma u_\sigma &=& u_\sigma = A \gamma_0 u_\sigma  + B u_\sigma = (A + B)u_\sigma \Rightarrow \nonumber \\
(A+B) &=& 1
\end{eqnarray}

By other hand, one gets:
\begin{eqnarray}
u_\sigma \bar{u}_\sigma u_\sigma \bar{u}_\sigma &=& (A \gamma_0 + B \openone)^2 = u_\sigma \bar{u}_\sigma = (A \gamma_0 + B \openone) \Rightarrow \\
A^2 + B^2 &=& B \nonumber \\
2AB &=& A
\end{eqnarray}
So one have a system of two equations plus
an identity, which is fullfilled by $A = B = 1/2$.
So one gets in the restframe:
\begin{eqnarray}
u_\sigma(0) \bar{u}_\sigma (0) &=& \frac{1}{2}\left( \gamma_0 + \openone \right)\\
v_\sigma(0) \bar{v}_\sigma (0) &=& \frac{1}{2}\left( \gamma_0 - \openone \right)\\
\end{eqnarray}
Finally, in order to generalize relation
to any frame with 3-momentum $p_k$ and energy $E$, 
one has to perform a Lorentz boost. It is easy to
see that from the equations below that
a Lorentz boost over $u_\sigma$ and 
$\bar{u}_\sigma$ spinors is given by 
\begin{eqnarray}
u_\sigma(p) &=&  \left( \begin{array}{cc} 
                  \exp \frac{1}{2}\beta_k \Sigma^k  & 0 \\ 
                  0 & \exp \frac{-1}{2}\beta_k \Sigma^k \end{array} \right)  u_\sigma(0) \label{boost} \\
\bar{u}_\sigma (p) &=& \bar{u}_\sigma (0) \left( \begin{array}{cc} 
                  \exp \frac{-1}{2}\beta_k \Sigma^k  & 0 \\ 
                  0 & \exp \frac{1}{2}\beta_k \Sigma^k \end{array} \right) \label{boostbar}
\end{eqnarray}
and they are a generalization for the equations (\ref{Ltrans}) for $\psi$, $\Omega$
spinors. Thus
\begin{eqnarray}
u_\sigma (p) \bar{u}_\sigma (p) &=& \left( \begin{array}{cc} 
                  \exp \frac{1}{2}\beta_k \Sigma^k  & 0 \\ 
                  0 & \exp \frac{-1}{2}\beta_k \Sigma^k \end{array} \right)  u_\sigma(0)
                  \bar{u}_\sigma (0) \left( \begin{array}{cc} 
                  \exp \frac{-1}{2}\beta_k \Sigma^k  & 0 \\ 
                  0 & \exp \frac{1}{2}\beta_k \Sigma^k \end{array} \right) \nonumber\\
           &=&    \frac{1}{2}\left\{ \left( \begin{array}{cc} 
                  \exp (\frac{1}{2}\beta_k \Sigma^k) \exp (\frac{-1}{2}\beta_l \Sigma^l) & 0 \\ 
                  0 & \exp (\frac{-1}{2}\beta_k \Sigma^k) \exp (\frac{1}{2}\beta_l \Sigma^l) 
                  \end{array} \right) \right.\nonumber \\
                  &&\;\;+\left.
                  \left( \begin{array}{cc} 
                  0 & \exp (\frac{1}{2}\beta_k \Sigma^k) \exp (\frac{1}{2}\beta_l \Sigma^l)   \\ 
                  \exp (\frac{-1}{2}\beta_k \Sigma^k) \exp (\frac{-1}{2}\beta_l \Sigma^l) & 0 \end{array} \right) \right\} \label{incmplt}
\end{eqnarray}
The way how this expression can be further simplified depends on the choice
of the representation $\Sigma_i$. Therefore we give the explicit
form for the cases discussed in this paper.
\begin{itemize}
\item
If the representation $\Sigma_i$ fulfills a Clifford algebra
\be\label{Clifford}
\{ \Sigma_i,\Sigma_j \}=2\delta_{ij}\openone\quad,
\ee
like it is the case for spin 
$\Sigma_i=\sigma_i$. One finds with $\beta_i=\beta n_i$
\begin{eqnarray}
u_\sigma (p) \bar{u}_\sigma (p) &=& \frac{1}{2}\left\{ \left( \begin{array}{cc} 
                  \openone & 0 \\ 
                  0 & \openone \end{array} \right) +
                  \left( \begin{array}{cc} 
                  0 & \openone\cosh (\beta)  +n_i \sigma^i \sinh (\beta)\\ 
                 \openone\cosh (\beta)  -n_i \sigma^i \sinh (\beta)& 0 \end{array} \right) \right\} 
                 \label{cmpltu}
\end{eqnarray}
and
\begin{eqnarray}
\nu_\sigma (p) \bar{\nu}_\sigma (p) &=& \frac{-1}{2}\left\{ \left(
\begin{array}{cc} 
                  \openone & 0 \\ 
                  0 & \openone \end{array} \right) -
                  \left( \begin{array}{cc} 
                  0 & \openone\cosh (\beta)  +n_i \sigma^i \sinh (\beta) \\ 
                 \openone\cosh (\beta)  -n_i \sigma^i \sinh (\beta)& 0 \end{array} \right) \right\} \label{cmpltv}
\end{eqnarray}
\item For the adjoint representation $i\epsilon_{klm}$, which is used for spin
one the corresponding algebra is
\be
\left\{
i \epsilon_i,i\epsilon_j
\right\}_{km}=2\delta_{ij}\openone_{km}-\delta_{im}\delta_{jk}
-\delta_{ik}\delta_{jm}\quad.
\ee
This implies that the exponential can be expanded as
\be
\exp (i\theta n_j \epsilon_j)_{km}=
\openone_{km}\cosh (\theta)+i n_j \epsilon_{jkm} \sinh (\theta)
+n_k n_m(1-\cosh (\theta))\quad.
\ee
With this the completeness relations for spin one read
\begin{eqnarray} \label{cmpltus1}
u_\sigma (p) \bar{u}_\sigma(p) |_{s=1}=
\frac{1}{2}\left\{ \left( \begin{array}{cc} 
                  \openone & 0 \\ 
                  0 & \openone \end{array} \right)\right.
\quad\quad\quad\quad\quad\quad
\quad\quad\quad\quad\quad\quad
\quad\quad\quad\quad\quad\quad
\quad\quad\quad\quad\quad\quad
\quad\quad\quad\quad\quad\quad \\ \nonumber+\left.
                  \left( \begin{array}{cc} 
                  0 & \openone_{km}\cosh (\beta)  
           +i n_j \epsilon_{jkm} \sinh(\beta)
                  -2 n_k n_m \sinh^2(\beta/2)\\ 
                 \openone_{km}\cosh (\beta)  
           -i n_j \epsilon_{jkm} \sinh(\beta)
                  -2 n_k n_m \sinh^2(\beta/2)& 0
\end{array} \right) \right\}
\end{eqnarray}
and
\begin{eqnarray}\label{cmpltvs1}
\nu_\sigma (p) \bar{\nu}_\sigma(p) |_{s=1}=
\frac{1}{2}\left\{ \left( \begin{array}{cc} 
                  \openone & 0 \\ 
                  0 & \openone \end{array} \right)\right.
\quad\quad\quad\quad\quad\quad
\quad\quad\quad\quad\quad\quad
\quad\quad\quad\quad\quad\quad
\quad\quad\quad\quad\quad\quad
\quad\quad\quad\quad\quad\quad \\ \nonumber-\left.
                  \left( \begin{array}{cc} 
                  0 & \openone_{km}\cosh (\beta)  
           +i n_j \epsilon_{jkm} \sinh(\beta)
                  -2 n_k n_m \sinh^2(\beta/2)\\ 
                 \openone_{km}\cosh (\beta)  
           -i n_j \epsilon_{jkm} \sinh(\beta)
                  -2 n_k n_m \sinh^2(\beta/2)& 0
\end{array} \right) \right\}
\end{eqnarray}
The relations  (\ref{cmpltus1}, \ref{cmpltvs1}) contain an
extra term $\sim n_k n_m$ which potentially causes problems
when causal propagation is studied. Whether
those problems can be solved or whether they 
are related to the known causality
problems in spin one field equations \cite{Velo:1970ur} remains
to be seen.
\end{itemize}
The completeness relations (\ref{cmpltu}, \ref{cmpltv})
are used when studying causality in the section \ref{spinstat}.


\begin{thebibliography}{15}

\bibitem{Dirac:1936tg}
  P.~A.~M.~Dirac,
  Proc.\ Roy.\ Soc.\ Lond.\  {\bf 155A}, 447-459 (1936).


\bibitem{Fierz:1939ix}
  M.~Fierz, W.~Pauli,
  Proc.\ Roy.\ Soc.\ Lond.\  {\bf A173}, 211-232 (1939).

\bibitem{Gelfand:1948}
 I.M.~Gel'fand and A.M.~Yaglom, Sov. Journ. JETP {\bf 18}, 703 (1948).

\bibitem{Bargmann:1948ck}
  V.~Bargmann and E.~P.~Wigner,
  Proc.\ Nat.\ Acad.\ Sci.\  {\bf 34}, 211 (1948).

\bibitem{Weinberg:1964cn}
  S.~Weinberg,
  Phys.\ Rev.\  {\bf 133}, B1318 (1964).

\bibitem{Chang:1967zz}
  S.~J.~Chang,
  Phys.\ Rev.\  {\bf 161}, 1308-1315 (1967).

\bibitem{Tung:1967zz}
  W.~K.~Tung,
  Phys.\ Rev.\  {\bf 156}, 1385 (1967).

\bibitem{Hagen:1970wn}
  C.~R.~Hagen, W.~J.~Hurley,
  Phys.\ Rev.\ Lett.\  {\bf 24}, 1381-1384 (1970).

\bibitem{Hurley:1971nz}
  W.~J.~Hurley,
  Phys.\ Rev.\  {\bf D3}, 2339-2347 (1971).

\bibitem{Hurley:1972ju}
  W.~J.~Hurley,
  Phys.\ Rev.\  {\bf D4}, 3605-3616 (1971).

\bibitem{Singh:1974qz}
  L.~P.~S.~Singh, C.~R.~Hagen,
  Phys.\ Rev.\  {\bf D9}, 898-909 (1974).

\bibitem{Singh:1974rc}
  L.~P.~S.~Singh, C.~R.~Hagen,
  Phys.\ Rev.\  {\bf D9}, 910-920 (1974).

\bibitem{Gershun:1979fb}
  V.~D.~Gershun and V.~I.~Tkach,
  JETP Lett.\  {\bf 29}, 288 (1979)
  [Pisma Zh.\ Eksp.\ Teor.\ Fiz.\  {\bf 29}, 320 (1979)].

\bibitem{Berends:1985xx}
  F.~A.~Berends, G.~J.~H.~Burgers, H.~van Dam,
  Nucl.\ Phys.\  {\bf B271}, 429 (1986).

\bibitem{Siegel:1986de} 
  W.~Siegel,
  Nucl.\ Phys.\ B {\bf 284}, 632 (1987).

\bibitem{Siegel:1986zi} 
  W.~Siegel and B.~Zwiebach,
  Nucl.\ Phys.\ B {\bf 282}, 125 (1987).


\bibitem{Howe:1988ft}
  P.~S.~Howe, S.~Penati, M.~Pernici, P.~K.~Townsend,
  Phys.\ Lett.\  {\bf B215}, 555 (1988).


\bibitem{Siegel:1988yz} 
  W.~Siegel,
  hep-th/0107094.

\bibitem{Berkovits:1996tn} 
  N.~Berkovits,
  Phys.\ Lett.\ B {\bf 388}, 743 (1996)
  [hep-th/9607070].


\bibitem{Metsaev:1997nj}
  R.~R.~Metsaev,
  [hep-th/9810231].

\bibitem{Francia:2002aa} 
  D.~Francia and A.~Sagnotti,
  Phys.\ Lett.\ B {\bf 543}, 303 (2002)
  [hep-th/0207002].

\bibitem{Niederle:2004bw}
  J.~Niederle, A.~G.~Nikitin,
  Phys.\ Rev.\  {\bf D64}, 125013 (2001).
  [hep-th/0412213].

\bibitem{Vasiliev:2004qz}
  M.~A.~Vasiliev,
  Fortsch.\ Phys.\  {\bf 52}, 702 (2004)
  [arXiv:hep-th/0401177].

\bibitem{Savvidy:2005vm} 
  G.~Savvidy,
  Fortsch.\ Phys.\  {\bf 54}, 472 (2006)
  [hep-th/0512012].

\bibitem{Bekaert:2006us} 
  X.~Bekaert, N.~Boulanger, S.~Cnockaert and S.~Leclercq,
  Fortsch.\ Phys.\  {\bf 54}, 282 (2006)
  [hep-th/0602092].


\bibitem{Francia:2007ee} 
  D.~Francia,
  Nucl.\ Phys.\ B {\bf 796}, 77 (2008)
  [arXiv:0710.5378 [hep-th]].

\bibitem{Francia:2008ac} 
  D.~Francia,
  Fortsch.\ Phys.\  {\bf 56}, 800 (2008)
  [arXiv:0804.2857 [hep-th]].

\bibitem{Engquist:2008rt} 
  J.~Engquist and O.~Hohm,
  Fortsch.\ Phys.\  {\bf 56}, 895 (2008)
  [arXiv:0804.2627 [hep-th]].

\bibitem{Campoleoni:2008jq} 
  A.~Campoleoni, D.~Francia, J.~Mourad and A.~Sagnotti,
  Nucl.\ Phys.\ B {\bf 815}, 289 (2009)
  [arXiv:0810.4350 [hep-th]].

\bibitem{Bengtsson:2009nk} 
  A.~K.~H.~Bengtsson,
  Fortsch.\ Phys.\  {\bf 57}, 499 (2009)
  [arXiv:0902.3915 [hep-th]].


\bibitem{Buchbinder:2009pa} 
  I.~L.~Buchbinder, V.~A.~Krykhtin and L.~L.~Ryskina,
  Nucl.\ Phys.\ B {\bf 819}, 453 (2009)
  [arXiv:0902.1471 [hep-th]].


\bibitem{Manvelyan:2010jr} 
  R.~Manvelyan, K.~Mkrtchyan and W.~Ruhl,
  Nucl.\ Phys.\ B {\bf 836}, 204 (2010)
  [arXiv:1003.2877 [hep-th]].


\bibitem{Campoleoni:2011hg} 
  A.~Campoleoni, S.~Fredenhagen and S.~Pfenninger,
  JHEP {\bf 1109}, 113 (2011)
  [arXiv:1107.0290 [hep-th]].

\bibitem{Polyakov:2011sm} 
  D.~Polyakov,
  Phys.\ Rev.\ D {\bf 84}, 126004 (2011)
  [arXiv:1106.1558 [hep-th]].

\bibitem{Chicherin:2011sm} 
  D.~Chicherin, S.~Derkachov, D.~Karakhanyan and R.~Kirschner,
  Nucl.\ Phys.\ B {\bf 854}, 393 (2012)
  [arXiv:1106.4991 [hep-th]].

\bibitem{Montero:2011za} 
  M.~Montero and E.~Martin-Martinez,
  Phys.\ Rev.\ A {\bf 84}, 012337 (2011)
  [arXiv:1105.0894 [quant-ph]].

\bibitem{Buchdahl:1958xv}
  H.~A.~Buchdahl,
  Nuovo Cim.\  {\bf 10}, 96-103 (1958).

\bibitem{Velo:1970ur}
  G.~Velo, D.~Zwanziger,
  Phys.\ Rev.\  {\bf 188}, 2218-2222 (1969).

\bibitem{Buchdahl:1982ni}
  H.~A.~Buchdahl,
  J.\ Phys.\ A {\bf A15}, 1057-1062 (1982).

\bibitem{Illge:1999tb}
  R.~Illge, R.~Schimming,
  Annalen Phys.\  {\bf 8}, 319-329 (1999).

\bibitem{Illge:1986vs}
  R.~Illge,
  Exp.\ Tech.\ Phys.\  {\bf 34}, 429-432 (1986).

\bibitem{Illge:1993cd}
  R.~Illge,
  Comm.\ Math.\ Phys.\  {\bf 158}, 433-457 (1993).

\bibitem{Deser:2001dt}
  S.~Deser and A.~Waldron,
  Nucl.\ Phys.\  B {\bf 631}, 369 (2002)
  [arXiv:hep-th/0112182].


\bibitem{Sorokin:2004ie}
  D.~Sorokin,
  AIP Conf.\ Proc.\  {\bf 767}, 172-202 (2005).
  [hep-th/0405069].

\bibitem{Zecca:2007ab}
  A. Zecca,
  Int.\ Jour.\ Th.\ Phys.\ {\bf 46}. 1045-1054. (2007).

\bibitem{Belinicher:1974am}
  V.~I.~Belinicher,
  Teor.\ Mat.\ Fiz.\  {\bf 20}, 320-337 (1974).

\bibitem{Green:1978dz}
  H.~S.~Green,
  Austral.\ J.\ Phys.\  {\bf 31}, 219 (1978).


\bibitem{Buchbinder:2007ak} 
  I.~L.~Buchbinder, A.~V.~Galajinsky and V.~A.~Krykhtin,
  Nucl.\ Phys.\ B {\bf 779}, 155 (2007)
  [hep-th/0702161].

\bibitem{Buchbinder:2008ss} 
  I.~L.~Buchbinder and A.~V.~Galajinsky,
  JHEP {\bf 0811}, 081 (2008)
  [arXiv:0810.2852 [hep-th]].


\bibitem{Belinfante:1953zz}
  F.~J.~Belinfante,
  Phys.\ Rev.\  {\bf 92}, 997-1001 (1953).



\bibitem{Brown:1958zz}
  L.~M.~Brown,
  Phys.\ Rev.\  {\bf 111}, 957 (1958).

\bibitem{Feynman:1958ty}
 R.P.~Feynman and M.~Gell-Mann,
 Phys. Rev. {\bf 109}, 193 (1958);

\bibitem{Proca:1936}
  A.~Proca, 
  Compt. Rend. 202, 1420 (1936).

\bibitem{Rarita:1941mf}
  W.~Rarita and J.~Schwinger,
  Phys.\ Rev.\  {\bf 60}, 61 (1941).


\bibitem{Koch:2011hc}
  B.~Koch and N.~Rojas,
  arXiv:1101.4619 [hep-th].



\end{thebibliography}
\end{document}